\documentclass[epj]{svjour}
\usepackage{graphics,amsmath,amssymb}
\begin{document}
\title{Relaxation phenomena at criticality}
\author{Andrea Gambassi\inst{1,2} 
}                     
%
%
\institute{Max-Planck-Institut f\"ur Metallforschung,
Heisenbergstr. 3, D-70569 Stuttgart, Germany.
\and 
Institut f\"ur Theoretische und Angewandte Physik,
Universit\"at Stuttgart,
Pfaffenwaldring 57, D-70569 Stuttgart, Germany.}
\date{Received: date / Revised version: date}
%
\abstract{%
The collective behaviour of statistical systems close to critical
points is characterized by an extremely slow dynamics which, in the
thermodynamic limit, eventually
prevents them from relaxing to an equilibrium state after a change in the
thermodynamic control parameters.
The non-equilibrium evolution following this change displays some of
the features typically observed in glassy materials, such as ageing,
and it can be monitored via dynamic susceptibilities and correlation
functions of the order parameter, the scaling behaviour of which is
characterized by universal exponents, scaling functions, and amplitude
ratios. This universality allows one to calculate these quantities
in suitable simplified models and field-theoretical
methods are a natural and viable approach for this analysis.
In addition, if a statistical system is spatially confined, universal
Casimir-like forces acting on the confining surfaces emerge and they
build up in time when the temperature of the system is tuned to its
critical value.
We review here some of the theoretical results that have been
obtained in recent years for universal quantities, such as the
fluctuation-dissipation ratio, associated with the non-equilibrium
critical dynamics, with particular focus on the Ising model with
Glauber dynamics in the bulk. The non-equilibrium dynamics of the
Casimir force acting in a film is discussed within the Gaussian
model.
} 
\maketitle
\newcommand\op{\varphi}
\newcommand\x{{\bf x}}
\newcommand\TR{t_R}
\newcommand\reff[1]{(\ref{#1})}
\newcommand\eq{{\rm EQ}}
\newcommand\beq{\begin{equation}}
\newcommand\eeq{\end{equation}}
\newcommand\kB{k_{\rm B}}
\newcommand\HH{{\mathcal H}}
\newcommand\FF{{\mathcal F}}
\section{Introduction}
\label{intro}
Critical points occur in the phase diagrams of a variety of
microscopically different systems, ranging from magnetic materials
(ferromagnetic/paramagnetic transition), to pure fluids (liquid/vapour critical
point), binary mixtures (mixing/demixing transition) 
and strongly interacting matter (QCD).
In spite of these differences at the microscopic scale, 
a unified picture of the collective critical
behaviour emerges in terms of the so-called {\it order parameter} $\op$ of the
transition, the nature of which depends on the specific system while its
fluctuations determine the physical behaviour close to critical points. In
particular,
in the case of ferromagnetic materials, $\op(\x,t)$ can be identified with the
coarse-grained magnetization density at point $\x$ and time $t$, whereas for
a fluid $A$/$B$ binary 
mixture with critical concentrations $c_A^{\rm crit}$/$c_B^{\rm
  crit}$, $\op$ is given alternatively, by $c_A-c_A^{\rm crit}$ or
$c_B-c_B^{\rm crit}$ where $c_{A,B}$ are the space-time dependent
concentrations of the species $A$ and $B$, respectively.
The thermal fluctuations of the order parameter field $\op$ are
correlated in space and time across a typical correlation length $\xi$ and
a (linear) 
relaxation time $\TR$. Upon approaching the critical point, $\xi$ and $\TR$
become much larger than the corresponding microscopic length and time scales
$\ell_{\rm micr}$ and $\tau_{\rm micr}$, respectively, resulting in a
spatially {\it collective} and temporally {\it slow} behaviour which is
characterized by a certain degree of {\it universality}. In fact, 
the physics at
scales much larger than $\tau_{\rm micr}$ and $\ell_{\rm micr}$ becomes
largely independent of the microscopic details of the system, 
depending only
on its gross features such as the range of the microscopic interactions, 
internal symmetries, spatial dimensionality $d$,
conservation laws etc. which characterize the so-called 
{\it universality class} of the transition. 
Universality is primarily an experimental fact based on the evidence 
that, e.g.,  
$\xi \sim \xi_0 |r|^{-\nu}$ and $\TR \sim \tau_0|r|^{-\nu
z}$ for $r\equiv (T-T_c)/T_c \rightarrow 0$, where
$T_c$ is the critical temperature, $\xi_0$ and $\tau_0$ specifically depend on
the microscopic details of the system and therefore are non-universal, whereas
$\nu$ and $z$ are universal exponents in the sense specified 
above~\cite{zj,HH}.
Universality can be exploited in order to 
study the critical behaviour of a statistical
system via suitable minimal models having the same gross feature as the
actual system and therefore belonging to the same universality class.
In most of the cases this is done in terms of effective theories for
the order parameter field $\op(\x,t)$ and the resulting critical scaling
properties  
can be analyzed by means of powerful field-theoretical and
renormalization-group methods~\cite{zj}.

The slow and collective behaviour which emerges close to critical points
affects relaxation phenomena and the way in which they actually occur. In
Sect.~\ref{sec:bulk} we discuss the case of 
{\it bulk} systems, with particular
emphasis on the emerging non-equilibrium properties, whereas in
Sect.~\ref{sec:conf} we focus on confined critical systems 
in film geometry.

\section{Non-equilibrium relaxation in the bulk}
\label{sec:bulk}

Altough the picture presented below is quite general, 
we shall refer specifically to the case of
uniaxial ferromagnets belonging to the Ising universality class,
with a critical temperature $T_c$ and a space-time dependent order
parameter $\op(\x,t)$ (magnetization density).  

\begin{figure}[h!]
\begin{center}
\resizebox{0.5\columnwidth}{!}{%
\includegraphics{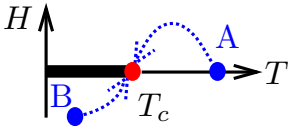}
}
\end{center}
\caption{Quenches in the phase diagram of an Ising ferromagnet: $T$ is the
  temperature of the bath, $H$ the external magnetic field and $(T,H)=(T_c,0)$
 is the critical point. The initial conditions A and B are characterized by a
  vanishing and non-vanishing magnetization, respectively.}
\label{fig:quench}       
\end{figure}

One of the simplest instance of relaxation phenomena 
can be observed after a sudden thermal perturbation. 
In fact, assume that a system, initially in 
equilibrium with a thermal 
bath at temperature $T_0$, is brought at time $t=0$ in
contact with a thermal bath at temperature $T\neq T_0$. As a consequence of
this perturbation the system starts evolving out of equilibrium in a way which
is expected to depend on the specific initial condition (e.g., on
$T_0$). After this transient regime, of typical duration $t_\eq(T)$, the
system equilibrates with the thermal bath at temperature $T$, 
loses memory of the thermal
quench and its dynamics becomes stationary and invariant (in the absence of
external fields) under time reversal; fluctuations are therefore described by
equilibrium dynamics.
The duration $t_\eq(T)$ of the transient regime, i.e., of the relaxation,
depends  on the temperature $T$ and is such that $t_\eq(T>T_c)$ is finite
whereas $t_\eq(T\le T_c)$ is actually infinite (in the thermodynamic limit)
and the non-equilibrium relaxation goes on forever. In particular, we focus
here on the case of a quench to the critical point $T=T_c$, referring the
reader to Ref.~\cite{Bray} for a discussion of the phase ordering dynamics for 
$T<T_c$. 
The fact that  $t_\eq(T=T_c)$ is not finite is actually due to 
the emergent slow and collective critical behaviour. In this context it is
important to understand %
the general features of this non-equilibrium relaxation,
exploring, e.g., how it depends on the initial condition of the
system. In particular -- referring to Ising-like magnets -- 
we will consider the cases depicted in fig.~\ref{fig:quench} of quenches from
an initial state with vanishing (A) and non-vanishing (B) magnetization. 
%
%
%
%
In addition to the dependence on the initial condition, it is also of interest
to understand how the universality which characterizes
critical phenomena manifests itself in the long-time relaxation and if a
scaling behaviour actually emerges. Some of these issues were already
addressed more than 30 years ago, with particular focus on the dynamics of the
order parameter~\cite{FR-76,BEJ-76}. Only quite recently, however, it has been
realized~\cite{CKP-94} that dynamic correlations and susceptibilities 
display very
interesting non-equilibrium properties such as those observed in glassy
systems, for example {\it ageing} (see, e.g., Ref.~\cite{C-02} for 
a review).  
In this respect, critical dynamics provides a valuable, non-trivial but
relatively simple instance of dynamics with ageing, which can be
characterized to an extent actually out of reach in more general cases of
glassy dynamics. 

\subsection{Dynamic observables}

\begin{figure}[h!]
\begin{center}
\resizebox{0.7\columnwidth}{!}{%
\includegraphics{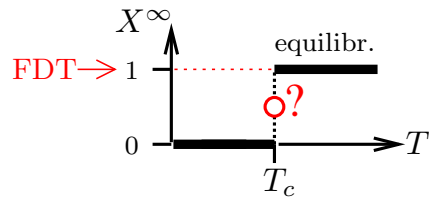}
}
\end{center}
\caption{$X^\infty$ as a function of the final temperature $T$ of the quench
  starting 
from a disordered initial state~\protect{\cite{GL}}. The value of $X^\infty$
for $T=T_c$ depends on the universality class of the system.}
\label{fig:Xinfty}       
\end{figure}

The relaxation process in a system of large volume $V$ 
can be monitored by looking at the global order
parameter $\Phi(t) = V^{-1} \sum_{x\in V} \op(x,t)$, its average over the
possible realization of the thermal noise  $M(t) = \langle\Phi(t)\rangle$ and
its two-time 
connected correlation function $C(t,s) = \langle\Phi(t)\Phi(s)\rangle_{\rm
  conn}$. In addition to $C(t,s)$ it is useful to analyse
the linear response
$R(t,s) = \delta M(t)/\delta H(s)|_{H=0}$ of the global order parameter $M(t)$
to a magnetic perturbation $H(s)$ applied at time $s<t$. With the aim of
highlighting non-equilibrium behaviours, we consider the
so-called {\it fluctuation-dissipation ratio} (originally introduced in the
study of glasses~\cite{FDR,CKP-94})
\beq
X(t,s) \equiv \frac{T R(t,s)}{\partial_s C(t,s)},
\label{eq:FDR}
\eeq
and in particular its limit for well-separated times
\beq
X^\infty \equiv \lim_{t\rightarrow\infty}\lim_{s\rightarrow\infty} X(t,s).
\label{eq:FDRinf}
\eeq
Indeed, for $t,s\gg t_\eq(T)$ the fluctuation-dissipation theorem (see, e.g.,
Ref.~\cite{C-02}) implies $X(t,s)=1$ and therefore $X^\infty=1$ whenever
equilibrium is eventually attained, which is always the case for $T>T_c$;
$X^\infty\neq 1$, instead, signals that the system evolves out of equilibrium
even at long times. In particular, for quenches from the disordered initial
state ($M(t<0)=0$) to $T<T_c$
and $H=0$ (domain coarsening), scaling arguments yield 
$X^\infty = 0$, leading to the global picture in
Fig.~\ref{fig:Xinfty}~\cite{GL}.
%
%
%
%
The value of $X^\infty$ for $T=T_c$, on the other hand, cannot be fixed on the
basis of general arguments but it depends on the specific system (actually on
the universality class which it belongs to~\cite{GL}). 
Note that, in contrast to the case of the non-equilibrium dynamics for $T<T_c$
(phase ordering) in which thermal fluctuations do not affect the leading
scaling behaviour~\cite{Bray}, the dynamics at $T=T_c$ is actually driven by
them and by the emerging collective behaviour of the system.
In recent years $X^\infty$ has attracted a lot of attention in connection
to the fact that, formally, it might be used to define an effective
non-equilibrium temperature $T_{\rm eff}$ for the system (see, e.g.,
Ref.~\cite{C-02} and references therein). 
Indeed, at large times, the rhs of Eq.~\reff{eq:FDR} equals 1 (as required by
the equilibrium fluctuation-dissipation theorem) if one assumes 
that the temperature
of the system is not given by the temperature $T$ of the bath but by $T_{\rm
  eff} = T/X^\infty$. In the case of mean-field glassy models it has been
shown that $T_{\rm eff}$ is actually a {\it bona fide} thermodynamic
temperature which controls, e.g., the direction of heat flows~\cite{C-02}.
In this context, the non-equilibrium critical relaxation provides a simple
instance of slow dynamics in which such a property can be tested beyond
mean field~\cite{cg-04}.

To give answers to the various questions posed so far, it is
possible to fully exploit universality~\cite{GL,cg-rev} in order to
characterize  within a field-theoretical 
approach~\cite{cg-02a1,cg-rev} the non-equilibrium critical relaxation and the
associated ageing phenomena 
(see Ref.~\cite{g-lec} for a pedagogical introduction).
This approach allows a systematic analysis of several aspects of these
relaxation phenomena and yields analytic predictions for scaling functions,
exponents, and amplitude ratios which characterize the scaling behaviour of
correlation and response functions at large times, within different
universality classes and dynamics. In addition, it is possible to highlight
dynamical crossovers in $C(t,s)$, $R(t,s)$~\cite{cg-co1,cg-co2} 
and in the persistence properties of $M(t)$~\cite{per}, 
due to different initial conditions.
%
%
Within this field-theoret\-ical approach,  
for example, instead of studying on a lattice 
${\mathbb Z}^d$
the non-equilibrium dynamics of the Ising model with spin-flip Glauber 
dynamics (which
captures the behaviour of some anisotropic ferromagnets and alloys), one
studies the Landau-Ginzburg 
{\it effective} Hamiltonian $\HH[\op]$~\cite{zj}  
for the order parameter $\op(\x,t)$ ($\x\in
{\mathbb R}^{d}$) with a suitable relaxational dynamics, known as Model
A~\cite{HH}:
\beq
\label{lang}
\partial_t \op(\x,t)=- D
\frac{\delta \HH[\op]}{\delta \op(\x,t)} +\zeta(\x,t).
\eeq
$D$ is a kinetic coefficient and
$\zeta(\x,t)$ a zero-mean stochastic Gaussian noise with 
$\langle \zeta(\x,t) \zeta(\x',t')\rangle= 2 \kB T D \, \delta({\bf x}-{\bf
  x}') \delta (t-t')$. 
The results presented below refer to this case.

\subsection{Scaling behaviour}

The renormalization-group analysis of Model A 
predicts that $R(t,s)$ and $C(t,s)$ display the following
scaling behaviours after a quench to $T=T_c$~\cite{JSS}:
\beq
s<t, \quad
\begin{cases}
R(t,s) = A_R t^{a\phantom{ + 1}}
{\displaystyle \left(\frac{t}{s}\right)^{\theta\phantom{-1}}} 
{\mathcal F}_R(s/t, t/t_0),
\\
C(t,s)  = A_C t^{a + 1}
{\displaystyle \left(\frac{t}{s}\right)^{\theta-1}}
{\mathcal F}_C(s/t,t/t_0),
\end{cases}
\label{eq:scaling}
\eeq
where $a=(2-\eta-z)/z$ is given in terms of well-known universal 
equilibrium static ($\eta$) and dynamic ($z$)
exponents~\cite{zj}, $\theta$ is the non-equilibrium universal 
initial-slip exponent~\cite{JSS}, ${\mathcal F}_{R,C}$ are universal scaling
functions normalized such that ${\mathcal F}_{R,C}(0,0)=1$. $A_{R,C}$ are
non-universal constants whereas $t_0$ is a non-universal time scale set by the
initial value of the magnetization $M_0\equiv M(t=0)$ and displays a
universal dependence on it
\beq
t_0 = B_m  M_0^{-1/\kappa}\,,
\label{eq:t0}
\eeq
where the universal scaling exponent $\kappa > 0$ is given, in terms of static
and dynamic equilibrium and
non-equili\-brium exponents, by $\kappa = \theta
+ a + \beta/(\nu z)$. The non-universal amplitude $B_m$ can be fixed by
suitable normalization conditions (see Refs.~\cite{JSS,cg-co1} for details).

Interestingly enough, Eqs.~\reff{eq:scaling} and~\reff{eq:t0}  indicate that
of the initial condition only the value $M_0$ of the magnetization really
matters in determining the scaling properties of the ensuing relaxation.
A more detailed analysis shows that correlations in the initial state 
(as long as they are short ranged) do only contribute to corrections to
the scaling behaviour~\cite{JSS}. 
According to Eq.~\reff{eq:scaling}, the two-time quantities $C(t,s)$ and
$R(t,s)$ are homogeneous functions of the three time scales $t$, $s$, and
$t_0$. In particular, when $s<t\ll t_0$, which is always the case if $M_0=0$,
the scaling form of $R$ (analogous one for $C$) becomes
\beq
R(t\ll t_0,s) = A_R t^a(t/s)^\theta f^{(0)}_R(s/t),
\label{eq:scalM0}
\eeq
where $f^{(0)}_R(x) = {\mathcal F}_R(x,0)$. Equation~\reff{eq:scalM0} 
clearly displays the scaling
behaviour typical of ageing phenomena: As a function of the time $t$ at which
the effect of the magnetic perturbation is measured,
the relaxation time is set by the time $s$ at which the perturbation was
applied, which is also referred to as the 
{\it age} of the system, being the time elapsed since the quench.
Upon increasing $s$ the ``response'' to the magnetic perturbation becomes
increasingly slow. The scaling behaviour in Eq.~\reff{eq:scalM0} was already
spelled out in Ref.~\cite{JSS} even tough the connection with ageing
has been realized some years later~\cite{CKP-94}. 
If $M_0=0$ the
scaling in Eq.~\reff{eq:scalM0} (and analogous for $C$) is valid also
at long times and allows one to express $X^\infty$ as an 
amplitude ratio 
$X^\infty = A_R/[(1-\theta)A_C]$~\cite{GL,cg-rev}. This amplitude
ratio turns out to be universal and therefore 
its value calculated within the field-theoretical
approach can be compared with the corresponding results obtained on the basis
of different
models in the same universality class (e.g., Ising model with Glauber
dynamics studied via Monte Carlo simulations).

In the opposite
limit of large times compared to $t_0$, i.e., $t_0\ll s < t$ the scaling form
of $R$  (analogous one for $C$) becomes~\cite{cg-co1}:
\beq
R(t,s \gg t_0) = a_R t^a(t/s)^{-\beta\delta/(\nu z)} f^{(\infty)}_R(s/t),
\label{eq:scalMinf}
\eeq
where $f_R^{(\infty)}(x) \sim x^{\theta+\beta\delta/(\nu z)} {\mathcal
  F}_R(x,y\rightarrow\infty)$ and $a_R$ is determined such that
$f_R^{(\infty)}(0) = 1$. 
Equation~\reff{eq:scalMinf} displays a scaling behaviour (with ageing) 
analogous to the 
one in Eq.~\reff{eq:scalM0}, with the major difference that the {\it
  non-equilibrium} exponent $\theta$ has been replaced by the combination
$-\beta\delta/(\nu z)$ of {\it equilibrium} static and dynamic 
exponents. Also in this case
$X^\infty$ can be expressed as an amplitude ratio $X^\infty =
a_R/\{[1+\beta\delta/(\nu z)] a_C\}$ where $a_C$ is the non-universal constant
in the scaling form of $C$ which corresponds to $a_R$.
According to
Eqs.~\reff{eq:scalM0} and~\reff{eq:scalMinf}, as soon as the initial value
$M_0$ of the magnetization is non-vanishing (quench from point B
in Fig.~\ref{fig:quench}), the asymptotic behaviour for
large times (and therefore the value of $X^\infty$) changes compared to the
case in which the initial value of the magnetization is zero (quench from
point A in Fig.~\ref{fig:quench}).  An analogous conclusion has been drawn for
different universality classes~\cite{co-varie,cg-co1,cg-co2}. The full scaling
functions in Eq.~\reff{eq:scaling} actually describe the crossover which
occurs, for a finite and fixed value of the magnetization $M_0$ (and therefore
$t_0$) when both times $s$ and $t$ increases from the case $s<t\ll t_0$
(described by Eq.~\reff{eq:scalM0}) to the
asymptotic case $t_0 \ll s < t$ (described by Eq.~\reff{eq:scalMinf}).

%
\begin{figure*}
\begin{center}
\begin{tabular}{ccc}
\resizebox{0.7\columnwidth}{!}{%
  \includegraphics{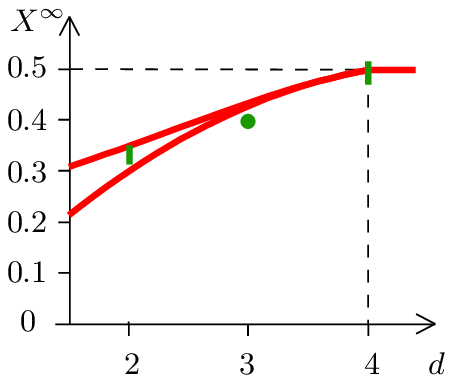}
}&\quad&
\resizebox{0.7\columnwidth}{!}{%
  \includegraphics{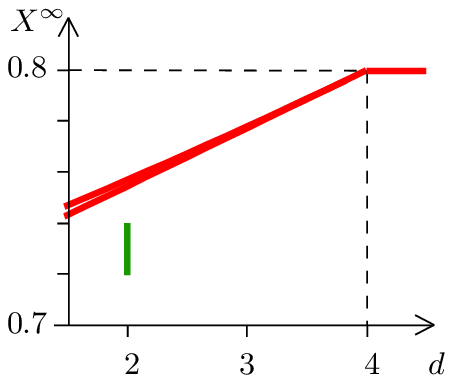}
}\\
(a)&&(b)
\end{tabular}
\end{center}
\caption{Asymptotic
  fluctuation-dissipation ratio $X^\infty$ (see Eqs.~\reff{eq:FDR}
  and~\reff{eq:FDRinf}) as a function of  
the space dimensionality $d$, within the Ising
  universality class with purely relaxational dynamics after a quench from a
  disorder (a) and magnetized (b) initial state. The solid lines correspond to
  different Pad\'e approximants of the two- (a) and one-loop (b) 
  series obtained
  via  the field-theoretical dimensional 
  expansion~\protect{\cite{cg-02a2,cg-co1}}. 
  For $d>4$ mean-field
  results become exact. The available numerical estimates obtained via Monte
  Carlo simulations~\protect{\cite{GL,MC-M0,cg-co1}}
  are also indicated with the corresponding errorbars, if
  provided. The actual agreement between analytical predictions and numerical
  results is quite good.}
\label{fig:XinfIsing}     
\end{figure*}
%
%
The field-theoretical approach yields not only the predictions, previously
discussed, for the scaling behaviour and the associated exponents but also
analytic expressions for other relevant universal quantities such as the
fluctuation-dissipation ratio and its asymptotic limit $X^\infty$. 
Here we focus on the specific case of the universality class of the
Ising model with Glauber dynamics, although analogous predictions are available
for different universality classes and dynamics (see Ref.~\cite{cg-rev} for
a review). 
In Fig.~\ref{fig:XinfIsing} we report the dependence of the asymptotic
fluctuation-dissipation ratio $X^\infty$ (see Eqs.~\reff{eq:FDR}
and~\reff{eq:FDRinf}) on the space dimensionality $d$ of the model, 
after a quench from an initial state with $M_0=0$ (a) and $M_0\neq 0$ (b) to
the critical point. The
solid lines correspond to different Pad\'e approximants of the $O(\epsilon^2)$
(a)~\cite{cg-02a2} and $O(\epsilon)$ (b)~\cite{cg-co1} 
series which have been calculated within the
field-theoretical $\epsilon$-expansion, where $\epsilon = 4 - d$. For $d>4$
the mean-field predictions $X^\infty(M_0=0)=1/2$ and $X^\infty(M_0\neq 0)=4/5$ 
become exact. %
The available Monte Carlo estimates in $d=2,3,4$~\cite{GL,MC-M0} for $M_0=0$ 
and in $d=2$ for $M_0\neq 0$~\cite{cg-co1},  
reported in Fig.~\ref{fig:XinfIsing}(a) and
(b), respectively, are in quite good quantitative 
agreement with the analytical predictions.

In summary, field-theoretical methods are a viable approach to investigate
non-equilibrium and ageing phenomena during critical relaxation in the bulk. 
The long-time properties of the non-equilibrium relaxation turn out to depend
on the initial condition (e.g., $T_0$) via the corresponding magnetization
$M_0$ and novel crossovers in the response, correlation functions and 
persistence properties of $M(t)$~\cite{per}  occur for
finite $M_0$, as confirmed by Monte Carlo simulations.
Within this approach one can also investigate the dependence of
$X^\infty$ -- and therefore of the effective temperature $T_{\rm eff} =
T/X^\infty$ -- on the observable used for its definition. Interestingly
enough, one finds~\cite{cg-04} that $X^\infty$ (equivalently, $T_{\rm eff}$) 
is independent of it only if
the fluctuations in the system are of Gaussian nature
(as in mean-field or spherical models). 
This observation might be of relevance to the case of
more complex glassy systems.

In addition to the case illustrated in this Section, a variety of different
universality classes and dynamics, 
both in the bulk and close to surfaces of semi-infinite systems have been
investigated in the literature~\cite{varia,cg-rev}. 
%


\section{Relaxation in confined geometries}
\label{sec:conf}

\begin{figure}[h!]
\begin{center}
\resizebox{0.45\columnwidth}{!}{%
\includegraphics{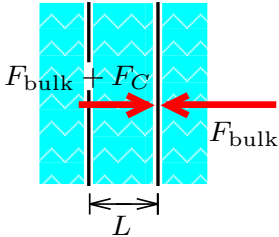}
}
\end{center}
\caption{Fluctuation-induced forces acting on two parallel walls immersed in a
  fluctuating medium. The net force on each wall is given by the Casimir
  term $F_C$.
}
\label{fig:Fwall}       
\end{figure}

\begin{figure}[h!]
\begin{center}
\begin{tabular}{cc}
\resizebox{0.5\columnwidth}{!}{%
  \includegraphics{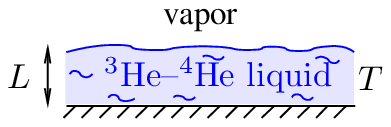}
}&
\resizebox{0.5\columnwidth}{!}{%
  \includegraphics{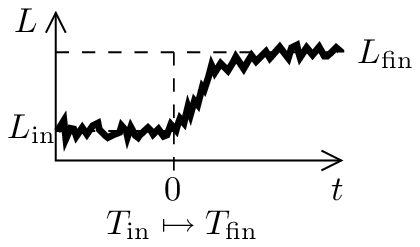}
}\\
(a)&(b)
\end{tabular}
\end{center}
\caption{(a) The equilibrium thickness $L$ of a wetting layer of a
  $^3$He/$^4$He mixture is affected by the Casimir force, as experimentally
  demonstrated in Ref.~\protect{\cite{gc}}. (b) After a sudden change of the 
  temperature from the initial value $T_{\rm in}$ to $T_{\rm fin}$, 
  the Casimir force is expected to change in time so that the film, originally
  of thickness $L_{\rm in}$, attains its final thickness $L_{\rm fin}$ after a
  relaxation.  
}
\label{fig:Hewe}       
\end{figure}

In Section~\ref{sec:bulk} we discussed some aspects of critical relaxation
in bulk systems, with particular focus on the non-equilibrium character of such
a dynamics and on quantitative predictions for the Ising universality class
with purely relaxational dynamics. In this Section we consider, instead,
relaxation
phenomena in confined geometries, focussing on 
{\it fluctuation-induced forces}
and on the way they build up in time after a thermal quench. Perhaps, the 
most widely known example of such forces is the Casimir effect
in quantum electrodynamics (QED)~\cite{Casimir};  
less known is the analogous effect which occurs in statistical physics,
discussed for the first time by Fisher and de Gennes~\cite{FdG}. Indeed,
whenever one confines a medium in which fluctuations (of different physical
nature) take place, effective forces arise on the confining walls. 
The medium can be constitued by, e.g., $^4$He, $^4$He/$^3$He
mixtures, classical binary mixtures, a Bose gas or even a magnetic material.
At variance
with QED, in statistical physics such effective forces 
have generically a finite range which is
related to the typical correlation length $\xi$ of the confined fluctuations.
When this correlation length becomes much larger than the typical microscopic
length scale $\ell_{\rm micr}$ in the system, the Casimir-like force $F_C$
acting on the confining
walls takes a {\it universal} form
\beq
\frac{F_C}{A} = \frac{\kB T}{L^d}\vartheta(L/\xi),
\eeq
where $A$ is the large transverse area of the $d-1$-dimensional 
walls, assumed to be parallel
and separated by a distance $L$.  $\vartheta$ is a {\it universal scaling
function} and therefore it is actually determined 
by the gross features of: (i) the system in the bulk and
of the fluctuating order parameter, as
mentioned in the Introduction, (ii) the gross features of the surfaces, which
-- to some extent -- are summarized by the boundary conditions they impose on
the order parameter $\op(\x,t)$ 
(see, e.g., Ref.~\cite{diehl} for a review) and (iii) the
geometry of the boundaries. In what follows we shall focus on the case of
parallel and infinite confining walls, i.e., on the film geometry. 
Due to universality, 
any minimal model with the same gross features as the confined systems we are
interested in captures its universal behaviour, including the Casimir force
$F_C$.
This model is usually 
specified in terms of an effective Hamiltonian $\HH[\op]$ which
determines the equilibrium distribution function $P_{\rm eq}[\op] \propto
\exp\{-\HH[\op]/(\kB T)\}$ of the
order parameter and therefore the effective free energy $\FF$ of the model. In
turn, close to the {\it bulk} critical temperature 
$T_c$ and for a system large enough 
(i.e., $L \gg \ell_{\rm micr}$), $\FF$ decomposes as the sum of 
a bulk term proportional to
the volume $V=A\times L$ of the system, 
one proportional to its surface area $A$ and a third
one which yields the leading and 
universal finite-size correction we are interested in:
\beq
\begin{split}
\FF = & \kB T_c A \times \\
& \left[ L f_{\rm bulk}(T) + f_{\rm surf}(T) +
  \frac{1}{L^{d-1}} \Theta(L/\xi) + \ldots \right],
\end{split}
\eeq
where higher-order terms in $L^{-(d-1)}\times (\ell_{\rm micr}/L)$ have been
neglected.
As a consequence of this decomposition, the force $F = - \delta \FF/\delta L$
acting on the confining walls 
is the  
sum of a bulk term $F_{\rm bulk}$ and the Casimir force
$F_C$ which is actually the force acting on the confining walls of the film
when they are ``immersed'' in the fluctuating medium (so that the contribution
$F_{\rm bulk}$ acting on the two opposite sides of each wall cancels, see
Fig.~\ref{fig:Fwall}).
%
%
%
%
Note that this way of determining the Casimir force $F_C$ via the equilibrium
free energy $\FF$ gives no information about the force at a given point in
space. 
A different approach which, in principle, 
provides such an information is based on the {\it stress tensor}
$T_{\alpha\beta}(\x)$. $T_{\alpha\beta}(\x)$ is a {\it local} functional of
the order parameter field $\op(\x)$, it is determined -- in equilibrium -- 
on the basis of $\HH$, and its expectation value on the equilibrium
distribution function $P_{\rm eq}$ yields the force  
\beq
F = \kB T A \langle
T_{\bot\bot}(\x)\rangle_{\rm eq},
\label{eq:Fstat}
\eeq
where $T_{\bot\bot}$ is a suitable
component of the stress tensor $T_{\alpha\beta}$ (see Ref.~\cite{krechbook} for
details). In equilibrium such an average is actually independent of the
position $\x$ at which it is calculated; nonetheless it is tempting to
interpret $T_{\bot\bot}(\x)$ as providing the {\it local fluctuating pressure}
(per $\kB T$) at the point $\x$ in space. 
Summing up, the equilibrium value of $F_C$ can be determined via both the free
energy $\FF$ and the stress tensor $T_{\alpha\beta}(\x)$. 
On the other hand, if one is interested in relaxation phenomena the definition
via $\FF$ is no
longer viable. Before discussing this point let us illustrate a simple
experimentally relevant setting 
in which the relaxation of $F_C$ might be observed. 
Indeed, consider the case in which the Casimir force, in competition with
dispersion forces, determines the equilibrium thickness $L$ 
of a wetting layer of a binary  $^3$He/$^4$He mixture at a given temperature
$T$ (see, e.g., Refs.~\cite{gc,mgd}), as in Fig.~\ref{fig:Hewe}(a).
%
%
%
%

%
\begin{figure*}
\begin{center}
\begin{tabular}{ccc}
\resizebox{0.95\columnwidth}{!}{%
  \includegraphics{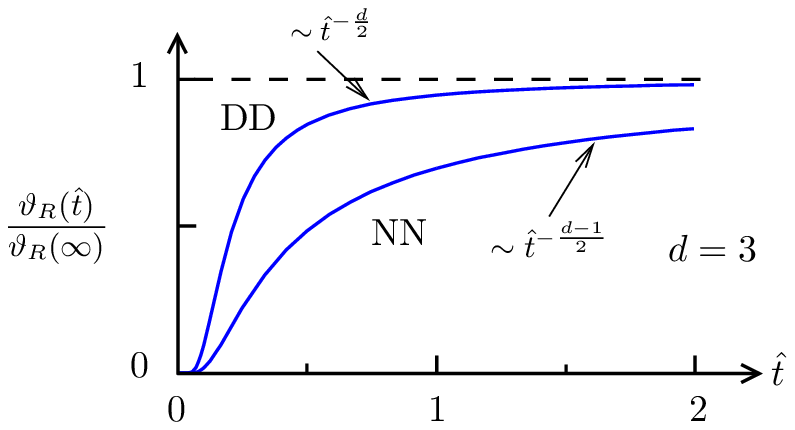}
}&\ &
\resizebox{0.95\columnwidth}{!}{%
  \includegraphics{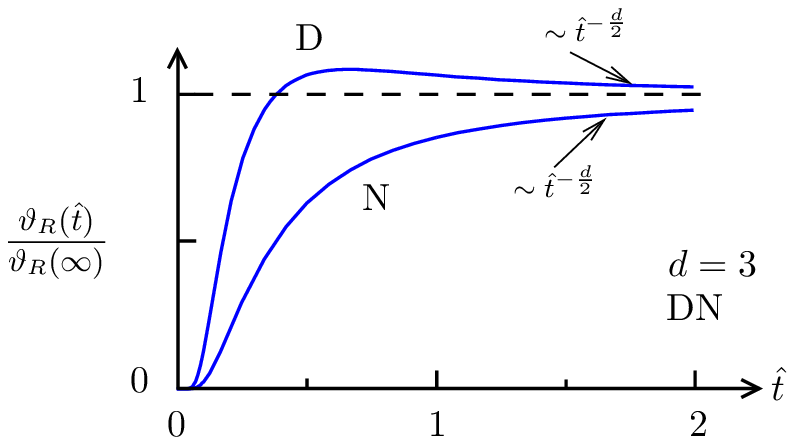}
}\\
(a)&&(b)
\end{tabular}
\end{center}
\caption{Universal scaling function $\vartheta_R(\hat t)$ of the Casimir force
  $F_C(t)$ in $d=3$ 
 after a quench from the high-temperature phase to the critical
  point (see Eq.~\protect{\reff{eq:FGauxRel}}) in the Gaussian model with
  purely relaxational dynamics. The finite-size scaling
  variable associated with time is $\hat t =
  (t/\tau_R)(\xi_0/L)^z$. The scaling function $\vartheta_R$ has been
  normalized by its asymptotic (equilibrium) value $\vartheta_R(\hat t
  \rightarrow \infty)$ (see Eq.~\protect{\reff{eq:FGaux}}). 
(a) $\vartheta_R(\hat t)$ for symmetric boundary conditions (DD and NN), 
leading to an attractive Casimir force which is equal on both the confining
walls. 
(b) $\vartheta_R(\hat t)$ for the case in which the walls
impose different boundary conditions on the order parameter (DN or ND). 
In contrast to equilibrium, the dynamic value and the qualitative features of
the time dependence of $F_C$ depend on the boundary condition imposed by the
wall at which  $F_C$ is measured. %
} 
\label{fig:CRel}       
\end{figure*}
%
%
%

Measuring $L$ as a function of $T$ allows a determination of the equilibrium
Casimir force. Upon changing suddenly the temperature (and therefore the
correlation length $\xi$ of the concentration fluctuations in the mixture) the
Casimir force $F_C$ changes and the thickness of the layer relaxes towards its
final value 
(as sketched in Fig.~\ref{fig:Hewe}(b)). Although experimental data for the
evolution of $L$ are not currently available, some aspects of the
dynamics of fluctuation-induced forces are expected to be accessible in the
near future. The evolution of $L$ is coupled to that one of the order
parameter field (and other relevant hydrodynamic quantities) which 
evolves in time under the effect of thermal fluctuations; in
turn, due to universality, this dynamics can be studied, close to critical
points, via simplified models (typically in the form of Langevin evolution
equations for the order parameter), as already mentioned in
Sect.~\ref{sec:bulk} (see Ref.~\cite{HH} for review).
Instead of discussing the problem within the 
experimental setup described above, in which $L$
changes in time, we assume that the fluctuating medium is confined between two
parallel walls at fixed distance $L$ while the force
$F_C(t)$ acting on them is measured (Fig.~\ref{fig:Fwall}). 
As in the case of static behaviour, the presence of the confining walls
imposes boundary conditions on the time-dependent order parameter field
$\op(\x,t)$, the dynamics of which is affected by the confinement. 
In order to determine the dynamics of the Casimir force one has to understand
how the local order parameter translates into the local force which can be
measured, say, at the walls. 
One natural connection between the two 
is provided by the local
stress which, via the time-dependence of the order parameter, becomes time
dependent: $T_{\alpha\beta}(\x,t) \equiv T_{\alpha\beta}|_{\op(\x) \mapsto
  \op(\x,t)}$. Accordingly, it is natural 
to {\it define} the dynamic force acting on the
walls $W$ as~\cite{gd-06} (compare with Eq.~\reff{eq:Fstat})
\beq
F(t) \equiv \kB T A \langle T_{\bot\bot}(\x\in W,t)\rangle_{\rm noise}
\label{eq:Fdyn}
\eeq
where the average $\langle\ldots\rangle_{\rm noise}$ is taken over the
possible realization of the thermal noise. 
Heuristically, this amounts to the assumption that at each time
there is an ``energy cost'' $\kB T A \delta L \langle
T_{\bot\bot}\rangle$ associated with the displacement $\delta L$ of one of the
confining wall, which is actually determined, as in equilibrium, by the
order parameter field.  
Note that in thermal equilibrium 
this definition renders the static force.
(See Ref.~\cite{ncas} for some proposed definitions of
fluctuation-induced forces in different non-equilibrium system.)
At variance with the equilibrium case, however, the local pressure
$F/A$ might actually depend on the point $\x$ at which it is measured. In what
follows we assume translational invariance in the direction parallel to the
walls, yielding a spatially constant $F$ on each of them. Finally, in order 
to obtain the
dynamic Casimir force $F_C(t)$ one has to subtract from $F(t)$ the
corresponding bulk contribution $F_{\rm bulk}(t) = \lim_{L\rightarrow\infty}
F(t)$. 
To see this definition at work we consider the case in which $T_{\rm in} \gg
T_c$, corresponding to a correlation length so small 
that each of the walls cannot feel the effect of the confinement imposed by
the other,
resulting in $F_C(t<0)=0$. For the final temperature, instead, we assume
$T_{\rm fin} = T_c$, which eventually leads to a non-vanishing, long-ranged
Casimir force $F_C(t>0)\neq 0$. (Note that, due to the critical-point shift in
the film geometry~\cite{NF}, 
$T_c$
{\it is not} the critical 
temperature 
of the film.) 
The analytic results illustrated below refer to the case in which the
confining walls impose Dirichlet (D: $\op(\x \in W,t)=0$) or Neumann
(N: $\partial_\bot\op(\x\in W,t) = 0$ where $\partial_\bot$ is the derivative in
the direction normal to the wall) boundary conditions onto the order
parameter field $\op$. 
We assume that the effective Hamiltonian $\HH$ which captures the equilibrium
properties of the system is quadratic in the order parameter (Gaussian model)
and that the dynamics is purely dissipative (Model A). (We point out, 
however, that the
actual dynamic of fluids and binary mixtures requires accounting for features
which are not present in Model A~\cite{HH}). 
As a consequence of the quadratic
Hamiltonian, the relevant component of the stress tensor can be expressed, in
terms of the order parameter field, as $T_{\bot\bot}(\x,t) = (\partial_\bot
\op(\x,t))^2/2 - (\nabla_\|\op(\x,t))^2/2$ where $\nabla_\|$ is the gradient
parallel to the wall. 

In the long-time limit the film attains equilibrium
and the force takes different values
depending on the combination of boundary conditions imposed by the two
confining walls:
\beq
\frac{F_C}{A} = \frac{\kB T_c}{L^d} \times
\left\{
\begin{array}{l}
\Delta^{\rm (s)}_d \ \,\, <0,\ \mbox{for DD, NN}, \\[1mm]
\Delta^{\rm (ns)}_d \ >0,\ \mbox{for DN, ND},
\end{array}
\right.
\label{eq:FGaux}
\eeq
where, in three dimensions $d=3$,  
the amplitudes for symmetric and non-symmetric boundary conditions are
$\Delta^{\rm (s)}_3 = -0.048\ldots$ 
and  $\Delta^{\rm (ns)}_3 = +0.036\ldots$, respectively (see, e.g.,
Ref.~\cite{krechbook}).
According to Eq.~\reff{eq:FGaux}, the Casimir force leads to {\it attraction}
between the confining walls if they impose the same boundary conditions on the
fluctuating order parameter (NN, DD), whereas {\it repulsion} results from
different boundary conditions (DN, ND). Note that in the latter case the
equilibrium force $F_C$ acting on the two walls is the same, independently of
the fact that they impose different boundary conditions on the field; 
in this sense they
cannot be distinguished by measuring $F_C$. 

In view of dynamical scaling, 
the relaxation of the force acting on the confining walls 
after the thermal quench is expected to 
depend on the temporal scaling variable 
$\hat t = (t/\tau_0)(\xi_0/L)^z$ via the dynamical scaling function
$\vartheta_R(\hat t)$, 
\beq
\frac{F_C(t)}{A} = \frac{\kB T_c}{L^d}\vartheta_R(\hat t =
\frac{t}{\tau_0}\left(\frac{\xi_0}{L}\right)^z),
\label{eq:FGauxRel}
\eeq
where $z=2$ is the dynamical exponent of the model we are focussing on. 
At small and large times we recover the initial vanishing force
and its final equilibrium value in Eq.~\reff{eq:FGaux}, respectively, yielding 
$\vartheta_R(0)=0$ and $\vartheta_R(\hat t \rightarrow \infty) = \Delta_d^{\rm
  (s,ns)}$, depending on the actual boundary conditions. 
In Fig.~\ref{fig:CRel} we report the analytic predictions for
$\vartheta_R(\hat t)$ (normalized to the corresponding equilibrium value
$\vartheta_R(\hat t \rightarrow \infty)$) 
for the case of symmetric DD, NN (a) and non-symmetric
DN, ND (b) boundary conditions. The approach to the asymptotic value 
depends on the boundary conditions but whereas in (a) it is independent of the
wall at which $F_C$ is measured (due to the symmetry), this is no longer the
case in (b). Accordingly, the dynamics of $F_C$ provides a way of determining
which one of the two walls imposes D or N boundary condition in the DN (or ND)
case, even though this is not possible by simply looking at the equilibrium
force. Interestingly enough, the time dependence is not monotonic in
the case of the D wall.

\begin{figure}
\begin{center}
\begin{tabular}{ccc}
\resizebox{0.38\columnwidth}{!}{%
  \includegraphics{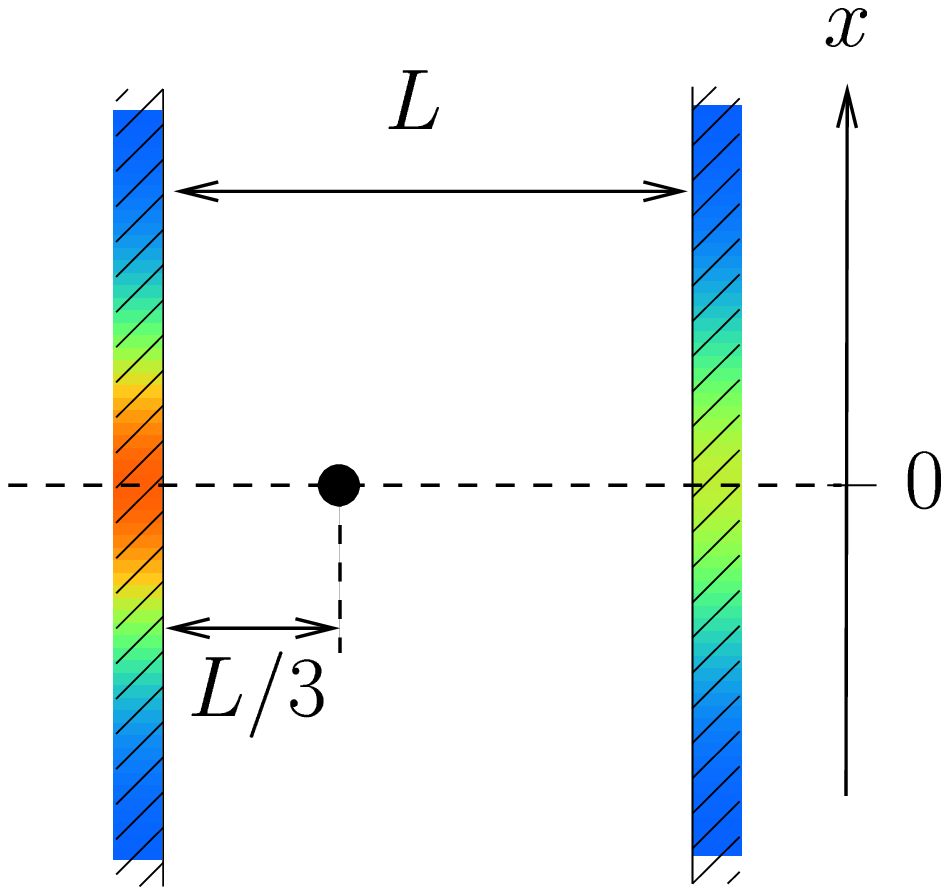}
}&\quad&
\resizebox{0.5\columnwidth}{!}{%
  \includegraphics{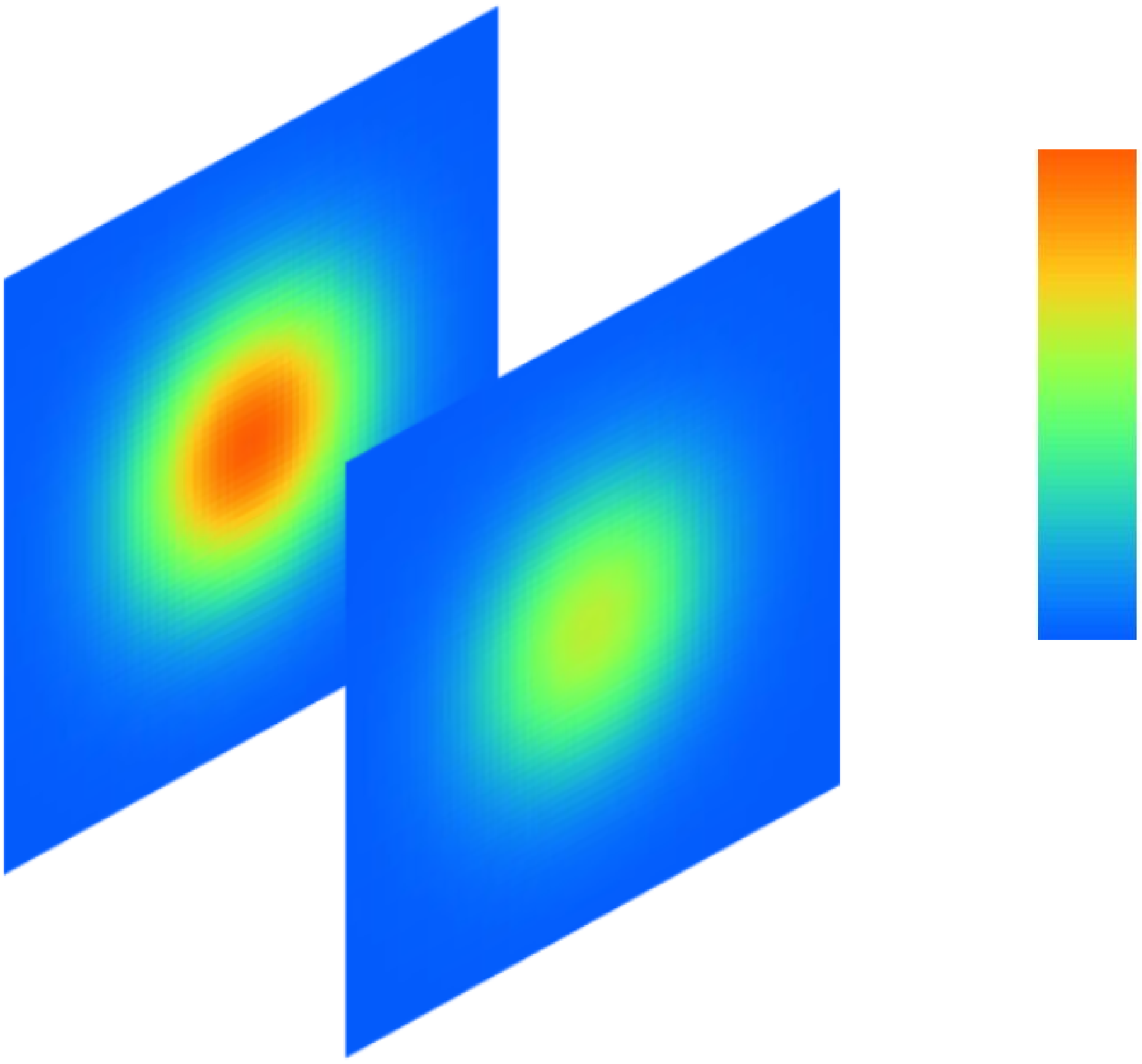}
}\\
(a)&&(b)
\end{tabular}
\end{center}
\caption{Effect of an external field on the Casimir force acting on the
  confining walls. The impulse field is applied at time $\hat t =0$ and
  distance $L/3$ from the left wall (a). The response of the medium is such
  that the  Casimir force locally increses compared to its (negative) 
  equilibrium value, reaches
  a maximum and then, due to the relaxational character of the dynamics,
  decreases again to its equilibrium value. The spatial dependence of the
  total local force at time $\hat t = 0.096$ for the critical Gaussian model
  with DD boundary conditions is sketched in (b), where the colorcode indicates
  increasing force from bottom to top, compared to the equilibrium value. The
  magnitude of the field-induced effect depends on the square of the strength
  of the applied perturbation (see Ref.~\protect{\cite{gd-06}} for details).  
}
\label{fig:extpert}       
\end{figure}

The definition~\reff{eq:Fdyn} allows one to study the effect of an external
perturbation on the Casimir force acting on the confining walls, as discussed
in detail in Ref.~\cite{gd-06} within the present model and DD 
boundary conditions. In particular, if the
external field (conjugate to the
order parameter) is localized in time and space, as indicated by the black dot
in  the side view in Fig.~\ref{fig:extpert}(a), the response starts to
propagate in the film, yielding asymmetric effects. 
At the very early stages, the response has practically
not yet reached the confining walls so that the force acting on them is
basically the equilibrium one corresponding to a vanishing external field. In
course of time the perturbation induced by the field hits the confining walls
and correspondigly the force exerted on them increases 
(reducing the equilibrium {\it attraction}),
with different magnitude at different points. Finally, because of the
relaxation character of the dynamics, the perturbation induced by the external
field vanishes in the limit of long times and the effective force reaches
again its equilibrium value. In Fig.~\ref{fig:extpert}(b) we report a sketch
of the distribution of the resulting stess on the confining wall at 
time $\hat t=0.096$ in the
case in which the impulse has been applied at time $\hat t = 0$ in the
position indicated in  Fig.~\ref{fig:extpert}(a). The system is assumed to be
at the bulk critical temperature $T_c$. Due to the different
distances from the two walls, the maximum force for $x=0$ is attained 
earlier on the left wall compared to the right one. (We  refer the reader to
Ref.~\cite{gd-06} for further details.)

%
%
%
%

In summary, we have proposed a definition of the {\it dynamic} Casimir force
$F_C$ via the local and time-dependent stress tensor $T_{\alpha\beta}$ (see
Eq.~\reff{eq:Fdyn}), which allows one to study the equilibrium and
non-equilibrium dynamics of $F_C$ in a variety of different cases, dynamics
and boundary conditions. The analysis of very simple models reveals already a
quite rich behaviour. However, it is desirable to establish a clearer
connection between this definition of the dynamic Casimir force and the force
that can be measured directly in actual experiments and molecular dynamics
simulations.

\section{Perspectives}

In the previous Sections we have reviewed some aspects of relaxation
phenomena in bulk and confined critical systems. 
Here we mention some issues which in our opinion deserve further
investigation. 

%
It is well known that surfaces affects locally both the static and dynamic
critical behaviour~\cite{diehl,p-rev} 
and therefore we expect that also non-equilibrium
properties such as the scaling behavior of two-time response and correlation
functions after a quench is accordingly changed. In spite of some available
Monte Carlo data and preliminary analyses~\cite{ag-surf,cg-rev}, 
a detailed and quantitative 
study within the field-theoretical approach is still lacking.
The slow collective relaxation phenomena discussed in Sect.~\ref{sec:bulk} are
due to the divergence of the equilibration time upon approaching an {\it
  equilibrium} critical point (characterized, e.g., by detailed balance). 
A qualitatively similar behaviour, however, has
to be expected also in the cases in which the system undergoes a
non-equilibrium phase transition of dynamical nature 
(e.g., reaction-diffusion systems, see
Ref.~\cite{H-07} and references therein). 
Also in this case, field-theoretical
methods provide valuable insight into the problem~\cite{bg-07} and it would be
desirable to extend the analysis to other relevant non-equilibrium
universality classes. 
A particularly intriguing question is how (and if) the non-equilibrium
critical relaxation phenomena in the bulk change when quantum fluctuations
come into play, e.g., upon approaching a quantum phase transition point.
%

%
The study of the dynamics of fluctuation-induced forces is still far from
being satisfactory and complete. In order to allow a comparison with
experimentally relevant settings (e.g., involving colloidal particles in
suspension) it is particularly important and urgent to explore the effects of
geometries and boundary conditions different from those discussed here 
and to consider dynamics which are more
suitable for the description of binary fluid mixtures, on which most of the
experiments are based. 

%
\begin{acknowledgement}
I am grateful to P.~Calabrese, S.~Dietrich, F.~Krzakala, A.~Macio\l ek,
R.~Paul and G.~Schehr for the stimulating collaborations which lead to some of
the results presented here.
\end{acknowledgement}
%
%
%
%

\end{document}